\documentstyle[psfig,aps,pra,twocolumn]{revtex}

\begin{document}

\title{Optical Nonlinear Dynamics with cold atoms in a cavity}

\author{A.Lambrecht, E.Giacobino, J.M.Courty}
\address{Laboratoire Kastler Brossel \thanks{Laboratoire de l'Universit\'{e}
Pierre et Marie Curie et de l'Ecole Normale Sup\'{e}rieure, associ\'{e} au CNRS}\\ 
Universit\'{e} Pierre et Marie Curie, case 74, 4 place Jussieu, 75252 Paris Cedex 
05, France}
\date{{\sc Optics Communications} {\bf 115}, p.199 (1995)}
\maketitle

\begin{abstract}
This paper presents the nonlinear dynamics 
of laser cooled and trapped cesium atoms placed inside an optical 
cavity and interacting with a probe light beam slightly detuned 
from the $6S_{1/2} F=4$ to $6P_{3/2} F=5$ transition. The system 
exhibits very strong bistability and instabilities. The origin 
of the latter is found to be a competition between optical pumping 
and non-linearities due to saturation of the optical transition.\\[2mm]
\end{abstract}

\section{Introduction}

In an atomic vapor, the interaction 
of the atoms with one or several near resonant electromagnetic 
fields is complicated by the fact that the various velocity classes 
have different detunings from the fields, except when this detuning 
is very large. Laser cooled atoms in a magneto-optic trap \cite{Raab87}
can have velocities as low as a centimeter per second. 
This means that their Doppler width is smaller than the natural 
linewidth and that a laser field can be set close to resonance 
with an atomic transition, all the atoms having the same detuning 
from the field. In such conditions, the interaction between the 
atoms and the field is well characterized and one can take advantage 
of the strong non-linearities of atomic systems while keeping 
the absorption rather small. In particular, it was shown that 
a probe beam going through a cloud of cold atoms could experience 
a strong gain due to Raman transitions involving the trapping 
beams \cite{Grison91}.
When the atoms are placed in a resonant optical cavity 
without a probe beam, laser action corresponding to that gain 
feature was demonstrated \cite{Hilico92}.

When cold atoms interact with a probe laser beam inside an optical 
cavity, bistability is easily observed at very low input powers 
(as low as $5 \mu$W) \cite{Giac94}. This bistability effect was observed in the 
presence of the cooling beams. However, to investigate the nonlinear dynamics 
of a collection of cold atoms in a cavity in more detail, we 
needed better controled conditions, and we studied the behaviour 
of the system in the absence of cooling beams, right after the 
trap is turned off. In that case, in addition to bistability, 
new features were found, such as very pronounced self pulsing 
oscillations in a wide range of experimental parameters.

While instabilities have been observed in similar conditions 
in atomic vapors \cite{Penna93}, the present situation is much easier to analyse, 
and we show hereafter that we have been able to find a rather simple 
model for these instabilities, that invokes a competition between 
optical pumping and saturation.

The exprimental procedure is described in section 2. In section 
3 we give a model explaining the dynamical behaviour of the system 
and compare its results with the measurements.

\section{Bistability and instabilities with cold atoms}

In the experiments described in the 
following, we prepare a cloud of cold cesium atoms in a cell 
using the background pressure to fill the trap. The trap operates 
in the standard way \cite{Raab87}, with three orthogonal circularly 
polarized trapping beams generated by a Ti:Sapphire laser and 
an inhomogeneous magnetic field. The Ti:Sapphire laser is detuned 
by 2.5 $\Gamma$ ($\Gamma$ being the linewidth of the upper state) 
on the low frequency side of the $6S_{1/2}F=4$ to $6P_{3/2}F=5$ transition. We 
obtain a cloud of cesium atoms 
the typical temperature of which is of the order of 1 mK, which 
gives a Doppler width much smaller than the natural width. The 
diameter (2.5 cm) and power (20~mW/cm$^{2}$) of our trapping 
beams allow us to obtain large clouds (about 5~mm in diameter) 
with densities of the order of 10$^{10}$~atoms/cm$^{3}$. The 
relevant parameter in the experiment is actually the number of 
trapped atoms in the probe beam, which is measured from the change 
in the intensity of the probe beam transmitted through the cavity 
with and without trapped atoms. This number is found to be ranging 
between 10$^{7}$ and 10$^{8}$ depending on the pressure of the 
background gas. As usual, the atoms non resonantly excited into 
the $6P_{3/2} F=4$ state and falling back into the $F=3$ ground 
state are repumped into the cooling cycle by a laser diode tuned 
to the $6S_{1/2} F=3$ to $6P_{3/2} F=4$ transition.

The cavity is a 25 cm long linear asymetrical 
cavity, close to half-confocal, with a waist of $260 \mu$m. Because the cell has 
optical quality 
antireflecting windows, we can build a good finesse optical cavity 
around the atomic cloud (Fig.1). Losses due to the two windows 
are of the order of 1\%. The input mirror has a transmission 
coefficient of 10\%, the end mirror is a highly reflecting mirror. 
The cavity is in the symmetry plane of the trap, making a 45${^\circ}$ angle with 
the two trapping beams that propagate in this plane.
\begin{figure}
\centerline{\psfig{figure=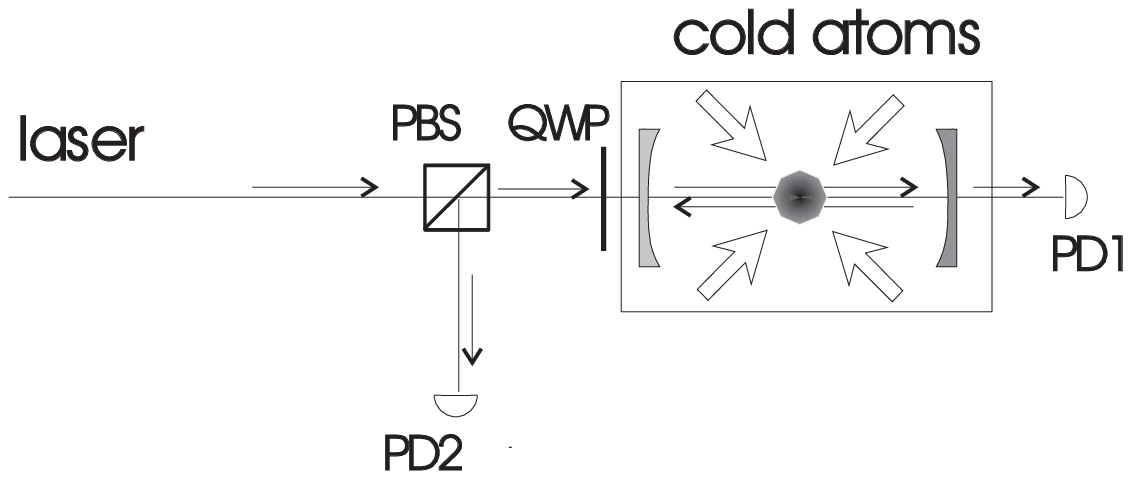,width=8cm}}
\caption{Experimental set-up showing 
the cell containing the cold atom cloud in an optical cavity; 
PBS: polarizing beamsplitter, QWP: quarter wave plate, PD1, PD2: 
photodiodes; PD1 and PD2 measure the powers respectively transmitted 
and reflected by the cavity.}
\end{figure}

To look for bistable behaviour of the optical cavity containing 
cold atoms, we send a circularly polarized probe beam into the cavity. It can be 
detuned by 0 to 130 MHz on either side of the $6S_{1/2} F=4$ to $6P_{3/2} F=5$ 
atomic transition.

We measure the power of the beam transmitted through the cavity 
while scanning the cavity length for a fixed value of the input 
intensity, as shown in Fig.2, or the input intensity for a fixed 
value of the detuning. The recording shows the characteristic 
hysteresis cycle due to bistability, where the output power switches 
abruptly between low and high values when the length of the cavity 
is scanned. Switching and hysteresis were also observed when 
the input power is scanned at fixed cavity length. In some cases 
overshoot and oscillations in the output power were recorded \cite{Giac94}.
\begin{figure}
\centerline{\psfig{figure=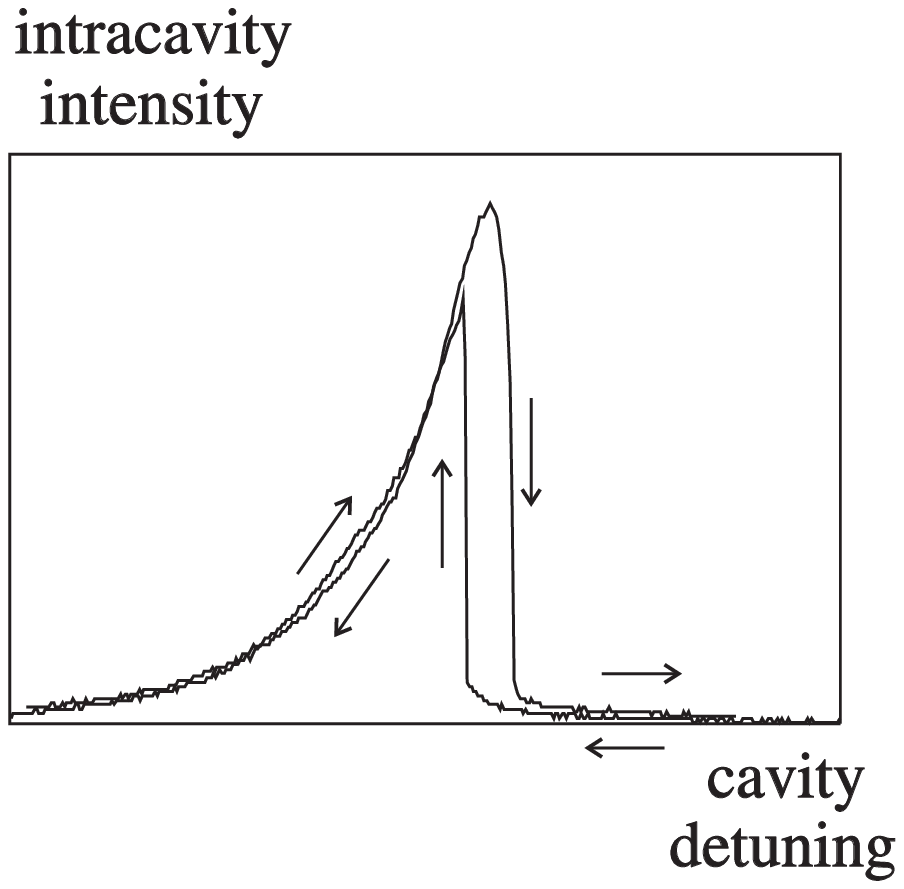,width=5cm}}
\caption{Recording showing the bistable switching from low to high 
transmission and back when the cavity length is scanned across 
the cavity resonance. The trapping laser beams are on. The laser 
is detuned by $22 \Gamma$ on the high frequency side of the 
atomic transition, the input power is 100~$\mu$W.}
\end{figure}

The shapes of the curves in such a case can be grossly interpreted 
from the theory of bistability with two-level atoms. However, 
the fit is approximate and the theory, even including single 
mode instabilities \cite{Boni78,Orozco89}, cannot explain the overshoot and 
oscillations \cite{Bram}. Actually the transition under investigation in cesium 
is far from being a two-level one in the presence of the trap. 
The fact that standard bistability is observed in most cases 
to be in fair agreement with two-level atom theory can be explained 
by the fact that the trapping beams randomize the the ground 
state population among the various Zeeman sublevels. 

To thoroughly investigate the phenomenon, it was thus indispensable 
to study the system without the trapping beams. After the trap 
is loaded, we turn off the trapping laser beams in order to get 
unperturbed atoms. We have about 20 ms to perform the measurements 
before most of the atoms have escaped out of the interaction 
region due to free fall and expansion of the cloud. In the experiment, 
the bistability parameter $C$ (see definition in section 3) can 
be as high as 400 just after the atoms have been released.

In a broad range of experimental parameters, we observed bistability 
and instabilities that exhibit unusal features. Instabilities 
are present within the whole range of accessible detunings (from 
$-25\Gamma$ to $25\Gamma$, where $\Gamma$ is the linewidth of the excited state, 
$\Gamma/2\pi=5.2$~MHz), and for input powers 
ranging from 50 to 300$\mu$W. These oscillations are somewhat 
similar to the ones observed in the presence of trapping beams 
but those were obtained only at high intensity or small atomic 
detuning. Here, on the contrary, they are observed very easily.

Fig. 3 shows a set of recordings of the output power of the 
cavity when the length is scanned, for different values of the 
input power. At very low power (not shown), the curves exhibit 
neither bistability nor instabilities. Starting for an input 
power of the order of $30\mu$W, oscillations appear on the left 
hand side of the cavity resonance curve, within some range of 
cavity detuning. This side is the one on which a bistable switching 
would occur in a saturated two-level atomic system. At higher 
input powers, the oscillations disappear and only bistability 
persists. Let us mention that when the power of the probe beam 
is too high, the atoms are expelled very rapidly from the beam 
\begin{figure}
\centerline{\psfig{figure=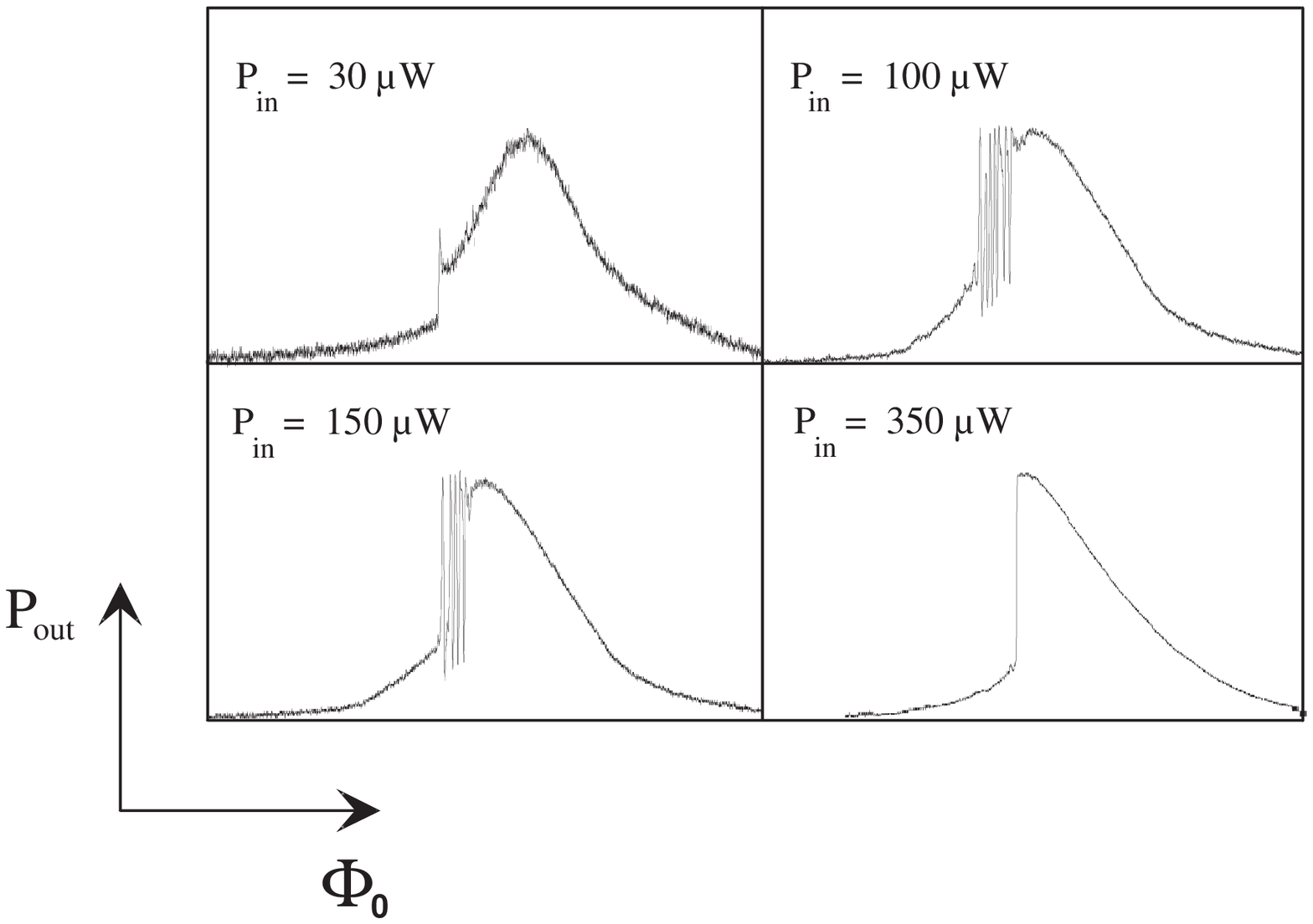,width=8cm}}
\caption{Recording of the output power P$_{\rm out }$of the cavity 
containing cold atoms when the cavity length is scanned, for 
four different values of the input power P$_{\rm in}$. The trapping 
beams are off. The atomic detunig is the same as in Fig. 2.}
\end{figure}
\noindent  
and only the very early part of the signal can be considered 
as significant of the nonlinear behaviour of the atoms.

Once enlarged these oscillations look clearly like self-pulsing. 
Their frequency is comprised between 100 KHz and a few MHz, and 
they are not due to the scan of the cavity length. To investigate 
them in more detail, one would want to record them at a fixed 
cavity length. However, it is not possible to keep the optical 
cavity length perfectly constant in time because the atoms escape 
from the original cloud, thus changing the index of refraction. 
But in such conditions, the effective cavity scan is slow and 
the oscillations are observed on longer time durations. Such 
a recording is shown in Fig.~4.
\begin{figure}
\centerline{\psfig{figure=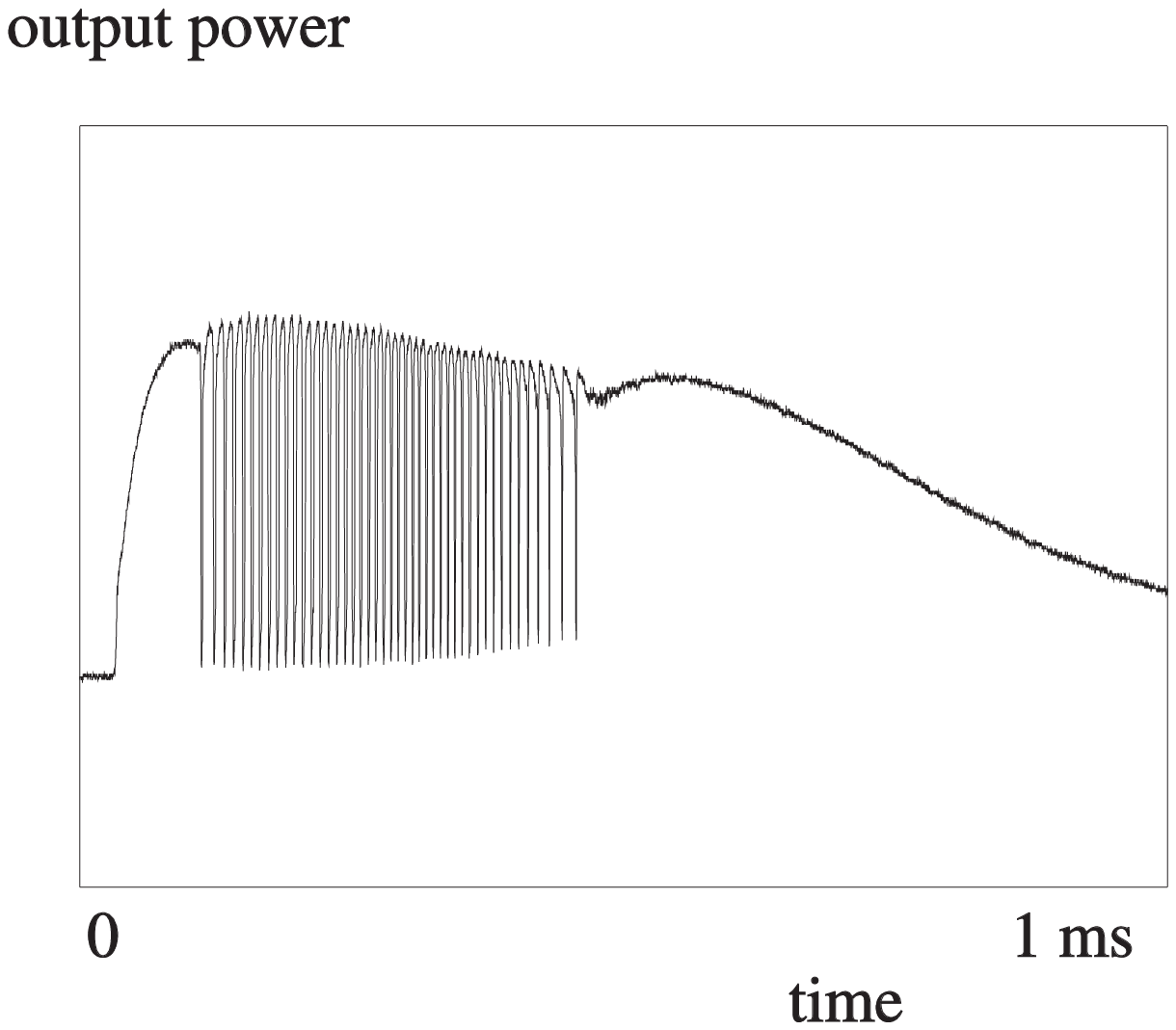,width=7cm}}
\caption{Recording of the instabilities for a fixed geometrical 
cavity length. The optical length is slowly scanned by the decrease 
of the number of atoms. The detuning is the same as in Fig. 2, 
the input power is 80~$\mu$W, the trapping beams are switched off at $t=0$.}
\end{figure}

\section{Model for instabilities}

To understand these oscillations, one 
has to take into account the hyperfine and Zeeman structure of 
the considered states. Various optical pumping effects can occur 
and phenomena linked to it like bistability, multistability 
\cite{Mlynek82,Cecchi82,Hamil83,Giac85} and instabilities 
\cite{Penna93,Boni78,Orozco89,Bram,Mlynek82,Cecchi82,Hamil83,Giac85,Gius87} have been 
predicted and observed in alkali vapors. Instabilities 
necessitate a strong coupling between the atoms and the field, 
that is a small detuning. Considering that the Doppler width 
is of the same order as the hyperfine structure in the ground 
and excited state of cesium, the system is rather intricate to 
describe, due to the competitive action of the various velocity 
classes and hyperfine sublevels. 

As already pointed out, the situation is much simpler with cold 
atoms, where one can consider that the field interacts with one 
hyperfine transition only. Owing to this, we have been able to 
develop a simple model to understand the origin of the observed 
oscillations. Roughly, as shown below, they result from the competition 
between a fast nonlinear process, the saturation of the optical 
transition and a much slower one, the optical pumping. 

The saturation of the optical transition causes a decrease of 
the linear index of refraction of the atomic medium when the 
intensity increases. On the contrary, optical pumping by circularly 
polarized light is at the origin of a non-linearity that increases 
the index of refraction with the light intensity: when the atoms 
are submitted to circularly polarized light, they tend to accumulate 
in the magnetic sublevels with high $m_{F}$ number, which have 
the largest coupling coefficient with the electromagnetic field. 
Therefore, the two nonlinear processes have opposite effects 
and compete in our system. The relaxation oscillations are a 
consequence of the significant difference in the characteristic 
times of theses processes.

Optical pumping tends to empty out the magnetic sublevels of 
low magnetic number to accumulate all the atoms in the sublevels 
with highest magnetic number ($m_{F}=4$ to $m_{F}=5$ transition). 
Due to the high number of magnetic substates, it takes a rather 
long time, starting from an equally distributed population, to 
complete the optical pumping to the highest $m_{F}$ sublevels 
of the ground and excited states. 

The time evolution of the sum of the populations of these two 
sublevels for several values of the Rabi frequency of the probe 
beam (calculated as an average over the hyperfine transitions) 
is shown in Fig. 5. One can see that even for values close to 
or larger than the saturation value, the optical pumping rate 
is much smaller than the natural linewidth. Thus the response 
time of the nonlinear susceptibility due to optical pumping is 
much larger than the one due to saturation of the optical transition, 
which, for a field detuned from resonance, can be considered 
as being of the order of the detuning.
\begin{figure}
\centerline{\psfig{figure=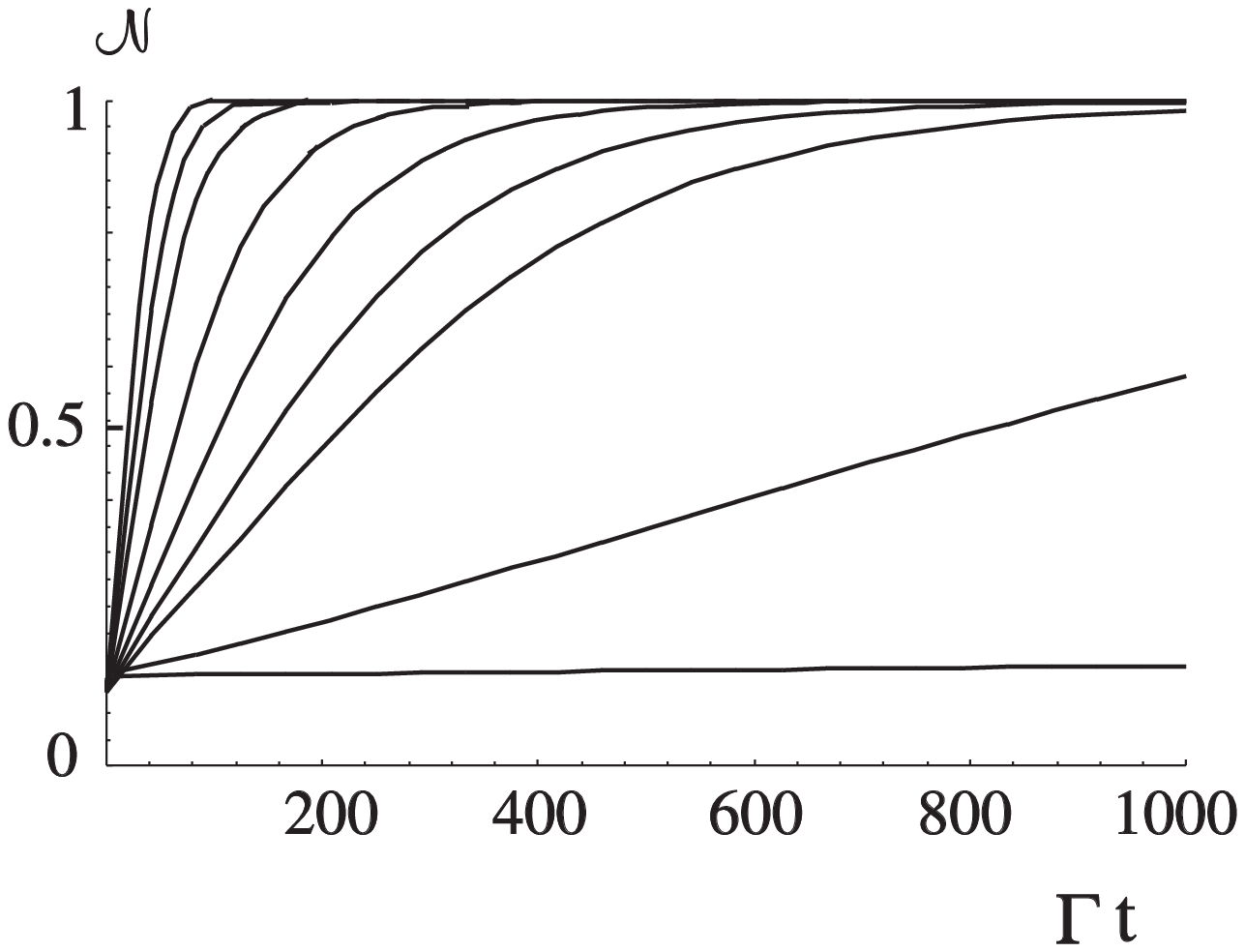,width=7cm}}
\caption{Optical pumping as a function of time : the curves show 
the calculated variation of the sum ${\cal N}$ of the populations of the $m_{F}=4$ 
sublevel of the ground state and of the $m_{F}$=5 sublevel of 
the excited state as a function of time, starting from a population 
equally distributed among the ground state sublevels, when the 
intensity I of the pumping light (defined by Eq. (1)) is equal 
to (from bottom to top)~: 1, 5, 10, 12, 15, 20, 30, 40, 60. Time is in units of 
$\Gamma$. The atomic detuning is $20 \Gamma$.}
\end{figure}

Although the behaviour of the system, involving many variables 
(all the hyperfine Zeeman populations and coherences and the 
field), is quite complex, the underlying mechanism can be explained 
with a simple model, involving basically two differential equations 
that give the evolution of the intracavity field and that of 
the atomic orientation in the ground state. To write these equations, 
we first introduce our basic notations.

The intracavity intensity $I$ of the laser is normalized to the saturation 
intensity :

\begin{equation}
I = \frac{g^2 |\alpha|^2}{\Gamma^2/4}, 
\label{int}
\end{equation}
where $g$ is the coupling constant of the atoms with the field,
\begin{equation}
g^2 = \frac{d^2 \omega_{\rm L}}{2 \epsilon_0 \hbar S c},
\label{g}
\end{equation}
$d$ is the atomic dipole, $\omega_{\rm L}$ the frequency of the probe laser and $S$ 
the cross section area of the beam.
$|\alpha|^2$ is the electric field squared, expressed in units of number 
of photons per second.

The round trip phase shift $\Phi_{\rm cav}$ of the field in the cavity 
(assumed to be a ring cavity) is the sum of four contributions. 
First, the phase shift $\Phi_{0}$ proportional to the geometrical length of the 
cavity. 
Second, we have two contributions due to the presence of atoms 
in the cavity, a linear phase shift,

\begin{equation}
\Phi_{\rm L} = 2Ng^2/\Gamma\delta
\label{PhiL}
\end{equation}
and a nonlinear Kerr-like phase shift
\begin{equation}
\Phi_{\rm NL} = -KI.
\label{PhiNL}
\end{equation}
The nonlinear coefficient K is given by
\begin{equation}
K = 4Ng^2/\Gamma \delta^3,
\label{K}
\end{equation}
where $N$ is the number of atoms and $\delta$ the detuning of the atomic transition 
frequency $\omega_0$ from the field frequency $\omega_{\rm L}$, normalized to the 
atomic transition 
linewidth $\Gamma/2$ 
\begin{equation}
\delta = 2(\omega_0 - \omega_{\rm L})/\Gamma.
\label{delta}
\end{equation}

$\Phi_{\rm L}$ and $\Phi_{\rm NL}$ are the phase shifts corresponding 
to the presence of two level atoms in the cavity. If we now consider 
that the ground and excited states have several Zeeman sublevels, 
the main additional contribution when the atoms interact with 
circularly polarized light is a term $\Phi_{p}$ coming from the change in the 
populations of the ground state sublevels and proportional to 
the ground state orientation $p$ (normalized to 1). Since the square 
of the Clebsch-Gordan coefficient of the $m_{F}$=4 to $m_{F}=5$ 
transition is about twice the mean square Clebsch-Gordan coefficient 
for the F=4 to F=5 transition, $\Phi_{p}$ is equal to $\Phi_{\rm L}$ for $p=1$, 
that is when the ground state is completely pumped. Thus, we can write
\begin{equation}
\Phi_p = \Phi_{\rm L}p.
\label{Phip} 
\end{equation}
The total phase shift in the cavity is then:
\begin{equation}
\Phi_{\rm cav} = \Phi_0 + \Phi_{\rm L} + \Phi_{\rm NL} + p\Phi_{\rm L}
\label{Phicav} 
\end{equation}

The ground state orientation $p$ increases with the intracavity intensity $I$ at 
rate $\beta I$ and decays at a rate $\gamma_{p}$ due to magnetic precession in 
transverse fields and to transitions to other hyperfine sublevels 
(via non resonant transitions):
\begin{equation}
{\rm d}p/{\rm d}t = -\gamma_p p + \beta I(1-p).
\label{pump} 
\end{equation}
The pumping rate~coefficient $\beta$ is computed from the calculation presented 
in Fig. 5 and the relaxation rate $\gamma_{p}$ is evaluated from the experimental 
parameters. 

The change of the intracavity field $\alpha$ on a round trip of time duration 
$\tau$ is due to the driving field $\alpha_{\rm in}$ entering through the coupling 
mirror of transmission {\it t}, to the mirror decay coefficient $\gamma_{\rm cav}$ 
(with $\gamma_{\rm cav}=t^2/2$) and to the round trip phase shift $\Phi_{\rm cav}$:
\begin{equation}
\tau {\rm d}\alpha/{\rm d}t = t\alpha_{\rm in} - (\gamma_{\rm cav} - i \Phi_{\rm 
cav})\alpha.
\label{champ} 
\end{equation}

The Kerr-like non-linearity has been 
assumed to have an instantaneous response. In the absence of 
optical pumping this system is well known to become bistable 
when the intensity is larger than a threshold intensity $I_{\rm bist}$ 
given by
\begin{equation}
I_{\rm bist} = 8\gamma_{\rm cav}^2/3\sqrt{3} K.
\label{bist} 
\end{equation}

In the presence of optical pumping 
eqs. (8) and (9) have to be solved numerically. They involve 
two different rates: the optical pumping rate $\beta I$ and the field evolution 
rate in the cavity $\gamma_{\rm cav}/\tau$ ($\gamma_{\rm cav}/2\pi \tau \approx 5$ 
MHz) which is usually larger than the optical pumping rate, except at very high 
pump powers.

These equations have been used to calculate the motion of the 
system. In some range of initial conditions, oscillations and 
limit cycles are found in the intracavity intensity as well as 
in the output intensity. To compare these results with the experimental 
data, we have calculated the output intensity when $\Phi_{0}$, i.e. the cavity 
length, is slowly scanned for several 
values of the input intensity. The result is shown in Fig. 6. 
It can be seen that the curves reproduce the experimental recordings 
in a satisfactory way. 

In particular, they show oscillations setting in for intermediate 
powers. When the input power is strong enough, the oscillations 
disappear, due to the fact that the optical pumping is fast enough 
to bring all the atoms in the sublevels of high magnetic number 
before the oscillations can start.  The calculations show that 
in the unstable region, the ground state orientation may vary 
by quantities as small as 1\%. As a consequence, the period of 
the instabilities can be much smaller than the typical optical 
pumping time shown in Fig. 5. 
\begin{figure}
\centerline{\psfig{figure=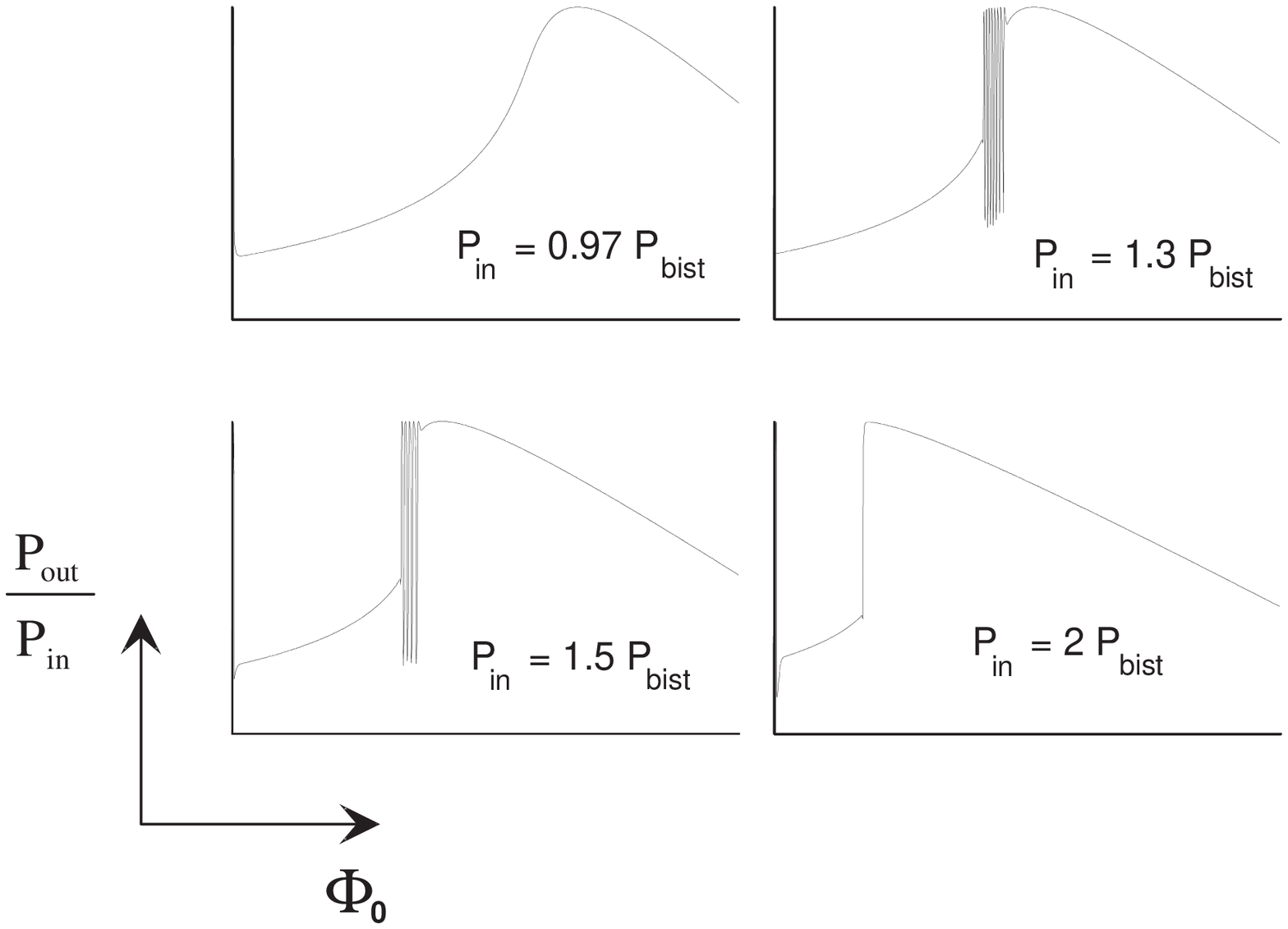,width=8cm}}
\caption{Calculated output power of the cavity containing cold 
atoms when the cavity length is scanned for four values of the 
input power chosen below and above the bistability threshold 
$P_{\rm bist}$ ($P_{\rm bist}$ is in the absence of optical pumping, 
cf eq.(10)).}
\end{figure}

This oscillatory behaviour can be understood by considering that 
the optical pumping changes the optical length of the cavity. 
It can then scan the cavity length back and forth in the vicinity 
of the bistable regime, causing periodic abrupt changes in the 
cavity transmission as shown in Fig. 7. A similar behaviour was 
observed in a different system with thermal effects \cite{Call78}.

Instabilities can also be found in the absence of relaxation 
for the orientation when the cavity length is scanned. This occurs 
for high intensities and large num- 
\begin{figure}
\centerline{\psfig{figure=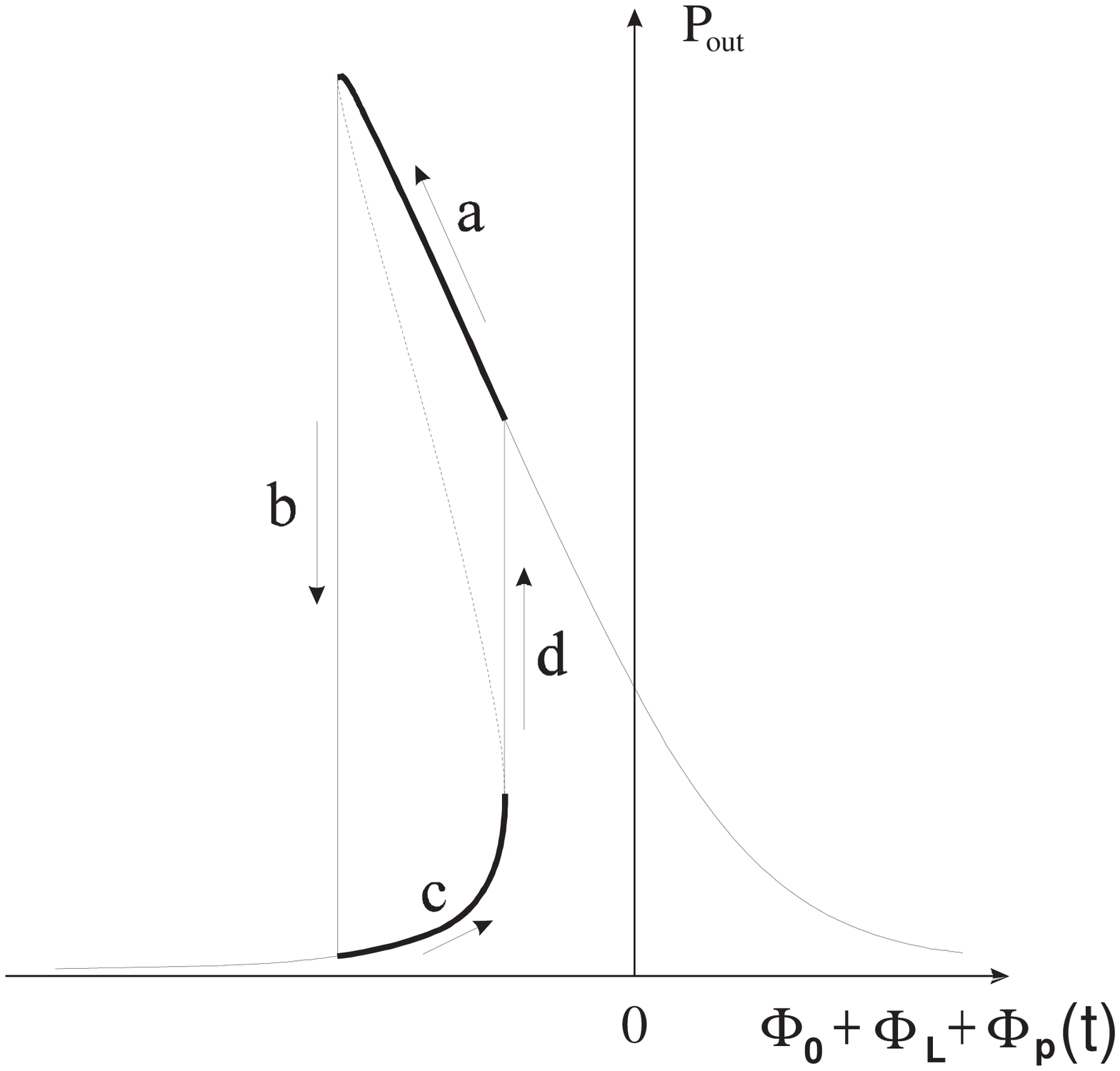,width=8cm}}
\caption{Diagramme showing the self-pulsing of the bistable system 
due to optical pumping (for the case of $\Phi_{\rm L}$ and $\Phi_{p}<0$); (a) the 
cavity is in the high transmission state and optical puming increases the 
orientation and decreases $\Phi_{\rm L}$; (b) abrupt switch towards the 
low transmission state; (c) the cavity is in the low transmission 
state, depumping is dominant and the orientation decreases; (d) 
switch towards the high transmission state.}
\end{figure}

\noindent bers of atoms. In such a case, 
in the vicinity of a resonance, the optical pumping takes over, 
brings the cavity first into resonance and then beyond the resonance. 
After this overshoot, the cavity length scan brings the cavity 
back into resonance, the optical pumping starts again and so 
on. Thus the orientation increases by steps, each time the light 
enters the cavity.

The validity of the model was checked by further experiments. 
When the atoms are released from the trap, it is possible to 
optically pump them into the $m_{F}=4$ magnetic sublevel of the 
ground state with an additional circularly polarized beam parallel 
to the probe beam, but closer to the atomic resonance. This prepumping 
is done in the presence of a magnetic field directed along the 
cavity. In such conditions the instabilities disappear. If the 
magnetic field is absent, the orientation created by the pump 
field is destroyed by the Larmor precession in tranverse magnetic 
fields and the instabilities persist.

A more complete treatment of the atomic non-linearity and of 
the optical pumping was also performed, where absorption and 
saturation of the optical non-linearity were taken into account. 
In this case, the separation of linear and nonlinear phase shifts 
is no longer possible. Instead, one defines a total phase shift 
due to two-level atoms
\begin{equation}
\Phi_1 = \frac{2Ng^2}{\Gamma} \frac{\delta+i}{1+\delta^2+2I}
\label{Phi1}
\end{equation}
or with the help of the bistability parameter $C$ given by 
\begin{eqnarray}
C &=& g^2N/\gamma_{\rm cav}\Gamma,\\
\Phi_1 &=& \frac{2C\gamma_{\rm cav}(\delta+i)}{1+\delta^2+2I}.
\end{eqnarray}

$\Phi_{1}$ now includes a contribution 
due to the linear absortion and dispersion of the atoms. The 
contributions $\Phi_{\rm L}$ and $\Phi_{\rm NL}$ introduced below are simply 
the first two real terms of the expansion of formula (11) in 
powers of I. The total phase shift, including optical pumping 
now writes
\begin{equation}
\Phi_{\rm cav} = \Phi_0 + \Phi_1 (1 + p).
\end{equation}

For consistency, the saturation of the optical pumping was also taken into account:
\begin{equation}
{\rm d}p/{\rm d}t = -\gamma_p p+\beta \frac{I}{1+\delta^2+2I}(1-p).
\end{equation}

The evolution of the system was calculated 
again using this more elaborate model. This yielded only minor 
changes in the results, the general behaviour predicted by the 
simple model for the instabilities remaining the same. 

Let us note that for high powers, the probe laser tends to push 
the atoms out of the beam, especially when its frequency is close 
to resonance. This phenomenon could give rise to bistability 
due to the change of the effective linear index of refraction 
of the medium with the number of atoms in the interaction zone. 
A careful study of the behaviour of the cold atoms in the probe 
beam has shown that such mechanical effects were negligible under 
our experimental conditions.

\section{Conclusion}

The recent development of the magneto-optic 
trap, which enables to get clouds of motionless atoms with a 
density comparable to that of the atomic beams is of great interest 
for nonlinear optics. In such traps, one can get large non-linearities 
by setting the driving field close to resonance without having 
much absorption. The instabilities described and interpreted 
in this paper show that it is possible to get new nonlinear phenomena 
in well characterized conditions. These experiments open the 
way to nonlinear and quantum optics using cold atoms in resonant 
cavities.
 
{\bf Acknowledgements:} The authors 
thank L.A. Lugiato for very useful comments and suggestions.
This work has been supported in 
part by the EC contract ESPRIT BRA 6934, the EC HMC contract 
CHRX-CT 930114 and the CNRS ULTIMATECH programme.

\end{document}